\documentclass[sigconf,table,xcdraw]{acmart}
\AtBeginDocument{%
  }
\renewcommand\footnotetextcopyrightpermission[1]{}
\usepackage{tabularx} 
\usepackage{makecell}  
\usepackage{multirow}
\usepackage{multicol}
\usepackage{etoolbox}
\usepackage{xcolor} 
\usepackage{tcolorbox}  
\usepackage{hhline}
\usepackage{pifont}      
\usepackage{colortbl}   
\usepackage{bbm}    
\usepackage{tikz}
\usepackage{stfloats}
\usepackage{float}
\usepackage{booktabs}

\definecolor{darkgreen}{rgb}{0.0, 0.5, 0.0}
\definecolor{tablegray}{RGB}{245, 245, 245}
\definecolor{titlegray}{RGB}{217, 217, 217}
\newcolumntype{Y}{>{\centering\arraybackslash}X}

\newcolumntype{C}[1]{>{\centering}m{#1}} 
\definecolor{topcolor}{RGB}{255, 220, 220}
\definecolor{botcolor}{RGB}{220, 235, 255}
\definecolor{lightblue}{RGB}{232, 240, 251}  

\newcommand{\topval}[1]{\textbf{\colorbox{topcolor}{\underline{#1}}}}
\newcommand{\botval}[1]{\textbf{\colorbox{botcolor}{#1}}}

\newcommand{\cmark}{\textcolor{green!70!black}{\ding{51}}} 
\newcommand{\xmark}{\textcolor{red}{\ding{55}}}             
\newcommand{\pmark}{%
    \begin{tikzpicture}[baseline=(A.base)]
        \node[inner sep=0pt, text=blue] (A) {\ding{51}};
        \draw[blue, line width=0.5pt, line cap=round] 
            ([xshift=-0.1pt, yshift=1.8pt]A.center) -- ([xshift=1.5pt, yshift=-1pt]A.center);
    \end{tikzpicture}%
}
\definecolor{rowblue}{RGB}{235, 235, 255}                  

\makeatletter
\patchcmd{\authornote}{\g@addto@macro\addresses{\@authornotemark}}{}{}{}
\makeatother

\begin{document}

\pagestyle{fancy}
\fancyhf{}                  

\fancyhead[RE]{\small\shortauthors}     
\fancyhead[LO]{\small\shorttitle}       

\title{RecRM-Bench: Benchmarking Multidimensional Reward Modeling for Agentic Recommender Systems}

\author{Wenwen Zeng$^{1,2^*}$, Jinhui Zhang$^{1,3^*}$, Hao Chen$^{1,4^*}$, Zhaoyu Hu$^{1\dagger}$, Yongqi Liang$^{1}$, Jiajun Chai$^{1}$, Dengcan Liu$^{1,5}$, Zhenfeng Liu$^{1}$, Shurui Yan$^{1}$, Minglong Xue$^{1}$, Xiaohan Wang$^{1}$, Wei Lin$^{1}$, Guojun Yin$^{1\ddagger}$}

\affiliation{
    \institution{$^{1}$Meituan, $^{2}$Fudan University, $^{3}$Nankai University}
    \institution{$^{4}$North China University of Technology, $^{5}$University of Science and Technology of China}
    \country{}
}
\email{wwzeng22@m.fudan.edu.cn, {huzhaoyu02,yinguojun02}@meituan.com}
\thanks{$^*$Equal contribution.}
\thanks{$^\dagger$Project leader.}
\thanks{$^\ddagger$Corresponding author.}

\renewcommand{\shortauthors}{Meituan-AsX Team}

\begin{abstract}
The integration of Large Language Model (LLM) agents is transforming recommender systems from simple query-item matching towards deeply personalized and interactive recommendations. Reinforcement Learning (RL) provides an essential framework for the optimization of these agents in recommendation tasks. However, current methodologies remain limited by a reliance on single dimensional outcome-based rewards that focus exclusively on final user interactions, overlooking critical intermediate capabilities, such as instruction following and complex intent understanding. Despite the necessity for designing multi-dimensional reward, the field lacks a standardized benchmark to facilitate this development. To bridge this gap, we introduce RecRM-Bench, the largest and most comprehensive benchmark to date for agentic recommender systems. It comprises over 1 million structured entries across four core evaluation dimensions: instruction following, factual consistency, query-item relevance, and fine-grained user behavior prediction. By supporting comprehensive assessment from syntactic compliance to complex intent grounding and preference modeling, RecRM-Bench provides a foundational dataset for training sophisticated reward models. Furthermore, we propose a systematic framework for the construction of multi-dimensional reward models and the integration of a hybrid reward function, establishing a robust foundation for developing reliable and highly capable agentic recommender systems. The complete RecRM-Bench dataset is publicly available at \url{https://huggingface.co/datasets/wwzeng/RecRM-Bench}.
\end{abstract}




\maketitle

\begin{figure}[t]
    \centering
    \includegraphics[width=0.5\textwidth]{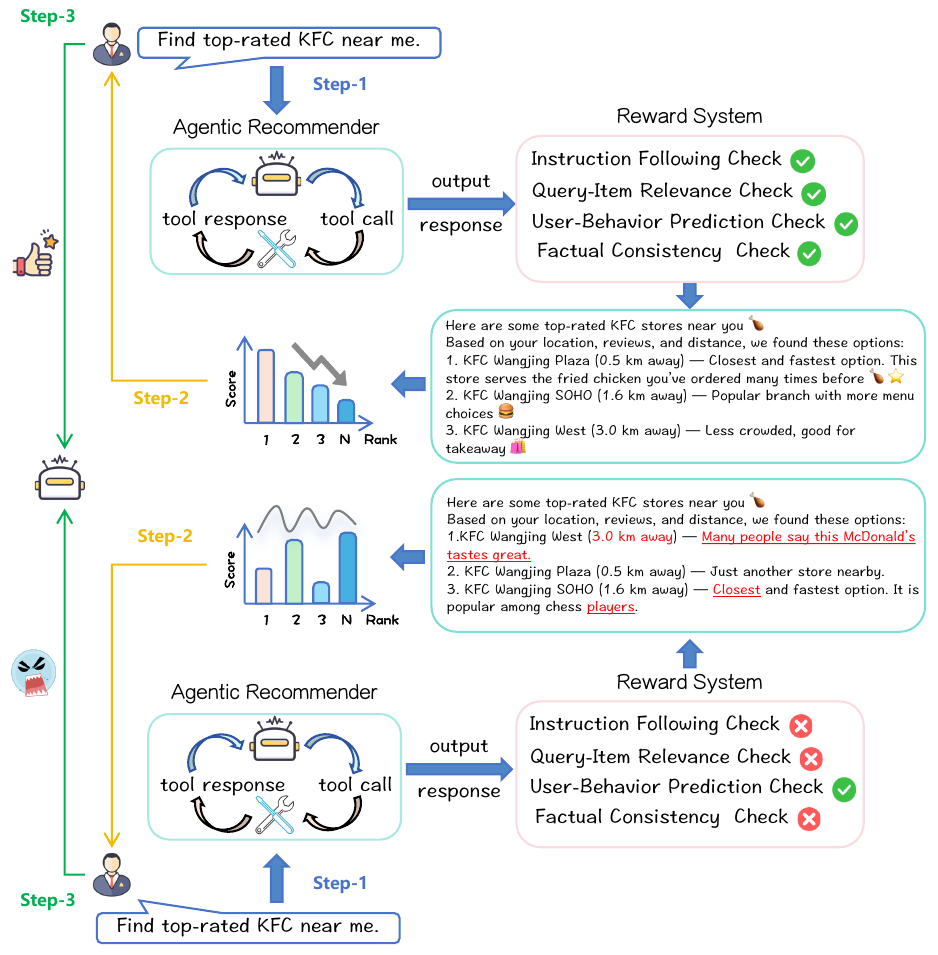}
    \caption{Advantages of reward systems built upon our proposed RecRM-Bench over existing methodologies in terms of multi-dimensional capability coverage.}
    \label{fig:fig_intro}
\end{figure}

\section{Introduction}
Recommender systems play a vital role in digital platforms such as e-commerce and social media by surfacing relevant items for users from vast information spaces ~\cite{zhang2019deep,zangerle2022evaluating,hussien2021recommendation,valencia2024artificial}. Traditionally, recommendation algorithms have centered on modeling user-item interactions to achieve personalized suggestions ~\cite{koren2009matrix,mnih2007probabilistic,koren2008factorization}. However, recent advancements in Large Language Models (LLMs) have introduced transformative possibilities for recommender systems ~\cite{chen2025toolforge}. By integrating LLM-powered agents, these systems can now deeply understand complex user intents, utilize external knowledge, and perform sophisticated reasoning and planning through tool integration ~\cite{wang2024recmind,he2025reindex,yu2025thought,huang2025mr,huang2025towards}. This enables a shift from simple query-item matching to highly personalized and context-aware recommendations.

To fully harness the generalization and reasoning capabilities of LLMs for recommendation tasks, Reinforcement Learning (RL) have been widely adopted to align agent behaviors with user preferences ~\cite{chen2023deep,wang2024reinforcement,wu2025starec}. However, existing research primarily focuses on optimizing overall user engagement or preference metrics, typically treating the reward signal as a single global outcome ~\cite{lin2025rec,xie2025recllm,zhang2025darlr}. Given that agentic systems involve lengthy, multi-step decision processes before generating a final recommendation, relying solely on terminal rewards leads to significant challenges. Specifically, the sparsity of user-item feedback results in unstable training and credit assignment problem ~\cite{li2025graphdrl}. Consequently, introducing explicit, process-level rewards is crucial for enabling stable and efficient learning in agentic recommender systems ~\cite{zheng2025deeprec,zhang2025process}.

Beyond the challenge of reward sparsity, existing research mainly focuses on single-dimensional reward signals, predominantly tied to final user actions such as clicks or ratings. Although user behavior prediction remains the core of recommendation systems, relying exclusively on interaction outcomes neglects other essential capabilities required for a reliable agentic recommender. A trustworthy agentic recommender should also adhere to rigorous operational integrity, such as ensuring syntactic compliance with output formats and reliably interacting with external tools to prevent system failure ~\cite{wang2025pro2guard,bi2025wemusic}. Furthermore, accurately interpreting user intent is also important for effective recommendation, as it ensures that the candidates retrieved from massive information spaces are semantically relevant to the user query, forming the logical prerequisite for effective interaction prediction ~\cite{xu2025enhancing} (comparisons are highlighted in Figure \ref{fig:fig_intro}). Optimizing solely for rewards related to user behavior fails to provide explicit training for these critical capabilities.

To support the development of agents that excel across these multiple dimensions, reward models that can assess instruction compliance, factual consistency, query-item relevance, and user behavior prediction are needed. However, the field currently lacks a unified benchmark for building and evaluating such multi-faceted reward mechanisms. The existing benchmarks mainly focus on behavioral prediction (as shown in Table ~\ref{tab:dataset_comparison}), while neglecting other critical dimensions required for robust recommendation performance.

To address these gaps, we introduce RecRM-Bench, the first comprehensive benchmark specifically engineered for reward modeling in agentic recommender systems. Our benchmark comprises a large-scale dataset of over 1.1 million entries, systematically covering four core evaluation dimensions in instruction following, factual consistency, query-item relevance, and user behavior prediction (includes specific item ranking and fine-grained behavior prediction). This provides a solid foundation for the comprehensive training and assessment of agentic recommender systems. Building upon RecRM-Bench, we further develop a standardized training paradigm and prompt framework for constructing multi-dimensional reward models. Our contributions are summarized as follows:
\begin{itemize}
    \item We construct the largest and most comprehensive benchmark for reward modeling in agentic recommender systems, supporting the holistic evaluation and development of multi-dimensional agent capabilities.
    \item We propose a systematic framework for reward model training based on RecRM-Bench, providing a robust data foundation for the development and optimization of agentic reward models.
    \item We introduce an integrative multi-dimensional reinforcement learning framework that leverages holistic feedback from our reward models, significantly reducing training variance and improving agent performance in complex, multi-step recommendation tasks.
\end{itemize}

\begin{table*}[t]
\centering
\caption{Comparison of existing user interaction benchmarks across key evaluation dimensions. ``\cmark'' indicates fully addressed, ``\protect\pmark'' indicates partially addressed, and ``\xmark'' indicates not addressed.}
\label{tab:dataset_comparison}
\small
\begin{tabular}{lcccccc}
\toprule
\multirow{2}{*}{\textbf{Benchmark}}
    & \multicolumn{2}{c}{\textbf{Agent Response-Related}} 
    & \multicolumn{3}{c}{\textbf{Recommendation-Related}} \\ 
\cmidrule(lr){2-3} \cmidrule(lr){4-6}
& \textbf{Instruction Following} & \textbf{Factual Consistency} 
& \textbf{Query-Item Relevance} & \textbf{Item Ranking} & \textbf{Behavior Prediction} \\ 
\midrule
JDSearch~\cite{jdsearch}                                        & \xmark & \xmark & \xmark & \cmark & \cmark \\
Qilin~\cite{qilin}                                              & \xmark & \xmark & \xmark & \cmark & \cmark \\
RecBench+~\cite{RecBench+}                                      & \xmark & \xmark & \pmark & \xmark & \cmark \\
AgenticShop~\cite{agenticshop}                                  & \xmark & \xmark & \pmark & \xmark & \cmark \\
RecIFBench~\cite{zhou2026openonerectechnicalreport}             & \pmark & \xmark & \xmark & \xmark & \cmark \\
KuaiSearch~\cite{li2026kuaisearchlargescaleecommercesearch}     & \xmark & \xmark & \cmark & \cmark & \cmark \\ 
\midrule
\rowcolor{rowblue} \textbf{\textsc{RecRM-Bench} (ours)}         & \cmark & \cmark & \cmark & \cmark & \cmark \\
\bottomrule
\end{tabular}
\end{table*}

\section{Related Work}
\subsection{Agentic Recommender Systems}
Recent advancements in Large Language Model (LLM) agents have introduced transformative paradigms for recommender systems. Current research generally follows three distinct trajectories based on the primary objective of the system ~\cite{peng2025survey}. Ranking centric agents utilize autonomous reasoning to infer user preferences directly from historical behavior ~\cite{wang2024recmind,wei2024llmrec,li2024large,xu2025iagent}, such as LLMRec ~\cite{wei2024llmrec}, and iAgent ~\cite{xu2025iagent}. In contrast, simulation centric agents ~\cite{zhang2024generative,wang2025user,bougie2025simuser,liu2025recoworld}, exemplified by Agent4Rec ~\cite{zhang2024generative} and SimUser ~\cite{bougie2025simuser}, leverage the role-playing capabilities of LLMs to emulate human like decision processes within simulated environments. Furthermore, interactive conversational agents ~\cite{10.1145/3726302.3729893,10572486,xu-etal-2025-beyond}, including the RAH framework ~\cite{10572486} and controllable dialogue simulators ~\cite{xu-etal-2025-beyond}, treat recommendation as a process of iterative intent refinement through multi turn interaction. Despite these contributions, existing methodologies often operate in isolation while neglecting the multi-dimensional capabilities required for a robust system.

\subsection{Benchmarking Recommender Systems}
Traditional recommendation benchmarks have primarily relied on domain-specific datasets rich in interaction data. MovieLens \cite{harper2015movielens} remains a cornerstone for rating-based evaluation, while the Amazon Review datasets \cite{mcauley2015image} provide diverse e-commerce scenarios across categories like Books and Beauty. Additionally, platforms such as Steam \cite{kang2018self} and Last.fm \cite{cantador2011second} offer large-scale interaction logs for the gaming and music industries. Despite their widespread adoption, these datasets focus almost exclusively on user-item interaction pairs, failing to capture the complex reasoning, tool integration, and multi-step decision-making inherent in agentic systems.

Recent research has begun tailoring benchmarks to the unique requirements of agents. However, most efforts still focus on the behavior prediction task. For example, AgentRecBench ~\cite{shang2025agentrecbench} establishes a benchmark for personalized systems through the simulation of user trajectories in interactive environments, while others \cite{jdsearch, qilin} emphasize intent recognition and hit rate prediction. With the development of agentic recommender systems, specific capabilities like instruction following and semantic relevance have gained attention. AgentIF \cite{qi2025agentif} evaluates adherence to functional constraints, and RecIF-Bench \cite{zhou2026openonerectechnicalreport} propose to examine instruction following, though the latter aligns more closely with intent recognition than with rigorous adherence to fine-grained constraints.  Even large-scale efforts like KuaiSearch \cite{li2026kuaisearchlargescaleecommercesearch}, which introduces ranking and relevance data, fail to offer a holistic perspective. These benchmarks typically address isolated components of agent behavior. In contrast, our work integrates instruction following, factual consistency, query item relevance, and user behavior prediction into a single framework, our work moves beyond simple interaction logs toward a systematic and reliable recommendation paradigm.

\section{Construction of RecRM-Bench}

\begin{figure*}
    \centering
    \includegraphics[width=\linewidth]{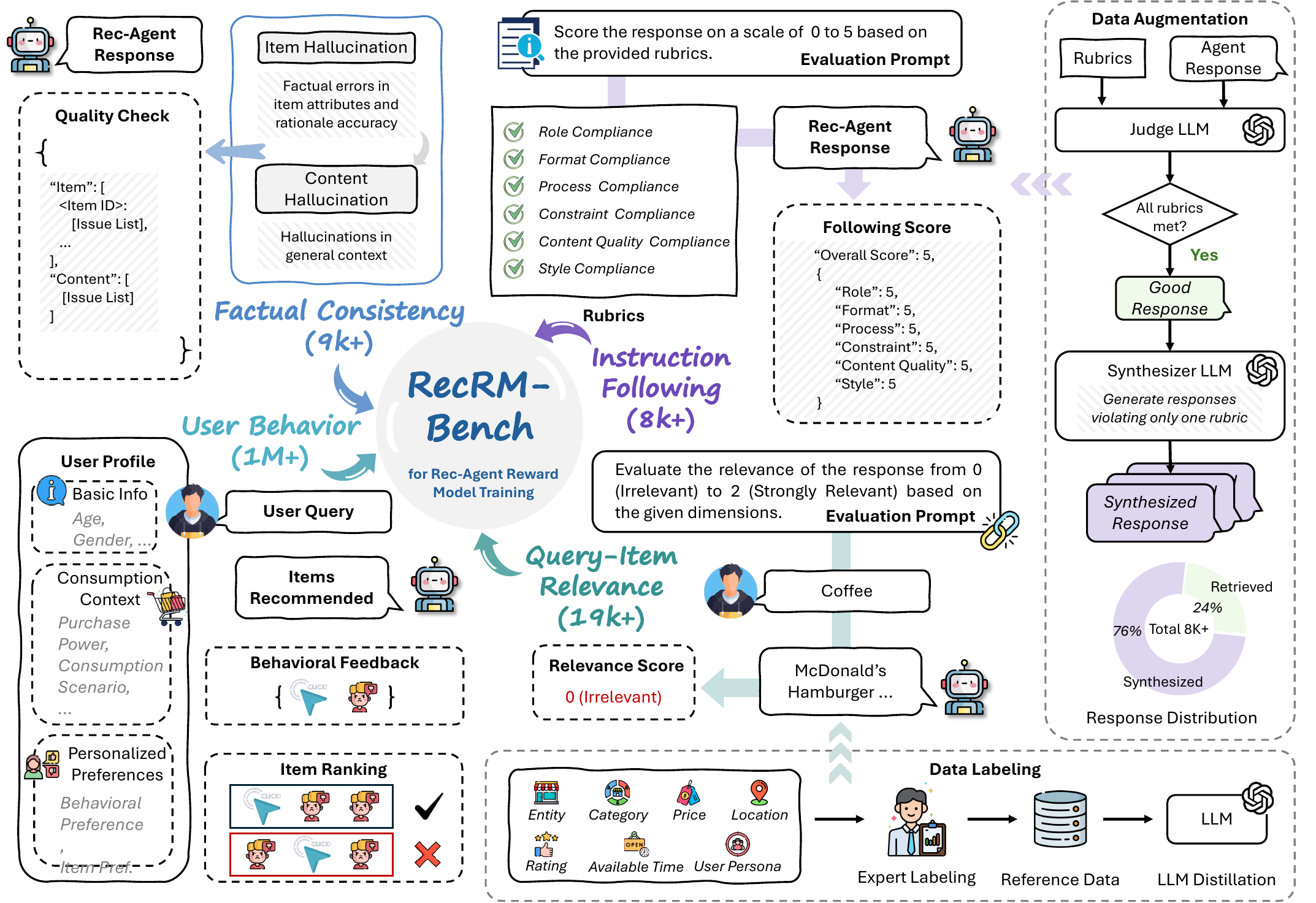}
    \caption{Overview of RecRM-Bench, encompassing four core evaluation dimensions: Instruction Following, Factual Consistency, Query-Item Relevance, and User Behavior Prediction.}
    \label{fig:benchmark_overview}
\end{figure*}

In this section, we introduce RecRM-Bench, a comprehensive benchmark derived from real-world interaction logs on platform Meituan. Each interaction is represented as ($u_i$,$q_i$,$r_i$), where $u_i$ is user-specific information, $q_i$ is the textual query issued by the user, and $r_i$ is the response of platform Meituan. The response $r_i$ includes a textual summary alongside a ranked list of recommended items $\mathcal{C}_i$. Based on these collected interactions, we construct four specialized databases tailored to evaluate specific dimensions of agentic performance, as illustrated in Figure ~\ref{fig:benchmark_overview}. The detailed construction of each database is as follows.

\subsection{Instruction Following Database Construction}\label{sec:section_if}
\subsubsection{Data Collection}
We collected a raw dataset comprising 68,096 query-response pairs from Meituan, a leading life-services platform. This dataset comprises 30,430 unique users and encompasses three distinct query types that reflect the complexity of real-world user intents: explicit-merchant (8.17\%), explicit-product (38.89\%), and multi-condition intent (52.94\%). These categories represent a spectrum of user needs, ranging from specific entity searches to complex, constraint-heavy requests. 

\subsubsection{Rubrics Generation}
To rigorously assess instruction-following fidelity, we established a set of comprehensive evaluation rubrics across six key dimensions: Role, Format, Process, Constraint, Content Quality, and Style Compliance. These rubrics were manually refined based on the system prompts and observed response patterns to capture the multi-dimensional nature of instruction adherence within recommendation-oriented tasks. Each dimension is further decomposed into fine-grained subfields (see Figure~\ref{fig:prompt_IF}) to ensure a standardized and objective scoring process.

\subsubsection{Data Augmentation and Synthesis}
Leveraging the expert-designed rubrics, we initially evaluated the raw responses using a dedicated LLM-as-a-judge (prompt shown in Figure ~\ref{fig:prompt_IF}). While the raw interaction data exhibited high diversity, we observed a long-tail distribution in compliance failures. To rectify this imbalance, we developed a targeted synthesis pipeline for data augmentation.

\begin{figure}
    \centering
    \includegraphics[width=\linewidth]{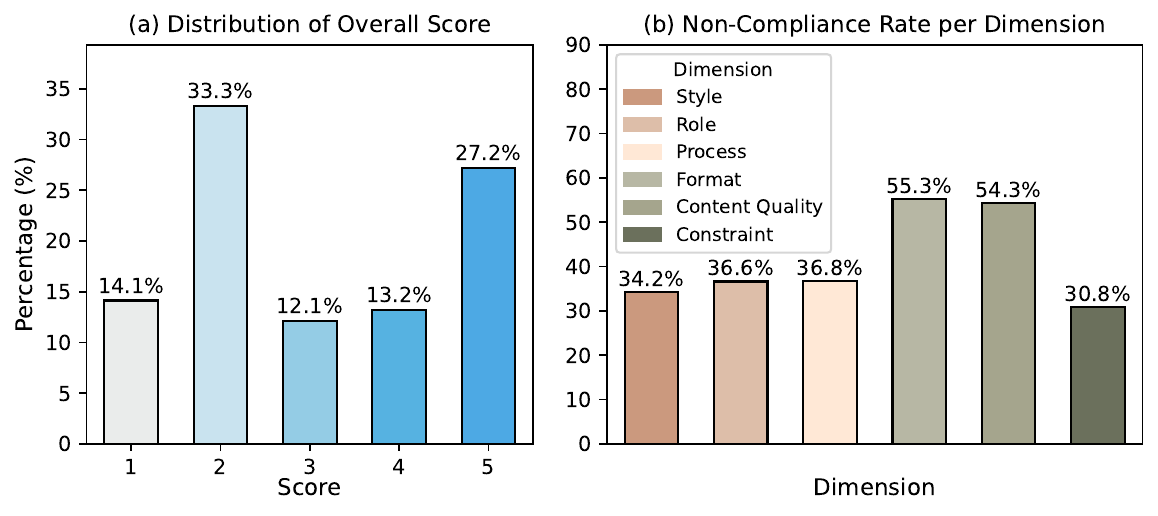}
    \caption{Statistical overview of overall scores and dimension-wise non-compliance in the Instruction Following database.}
    \label{fig:IF_score_bar}
\end{figure}

We first employed rejection sampling to identify 2,000 seed instances that achieved perfect compliance across all rubrics. From these seeds, we systematically generated negative samples by applying a controlled synthesis prompt (see Figure~\ref{fig:prompt_IF_syn}) to introduce a single, isolated violation for a specific dimension while keeping others intact. This counterfactual synthesis yielded 6,422 high-quality negative samples. By integrating these with the positive seeds, we constructed a balanced training set for the instruction-following reward model. The resulting dataset comprises query-response pairs labeled with fine-grained, multi-dimensional compliance scores, with the detailed score distribution summarized in Figure~\ref{fig:IF_score_bar}.

\subsection{Factual Consistency Database Construction}
Although hallucination is included as an evaluation dimension in our Instruction Following Database, experimental results show that standard instruction checks fail to cover all hallucination issues. Since reliability is one of the important requirements for agentic recommender systems, we propose a hallucination-specific database for reward modeling to provide better guidance on factual consistency.

\subsubsection{Data Collection} Based on our error analysis, we identify and categorize hallucinations into two primary types: Item Hallucination and Content Hallucination. Item Hallucination refers to the fabrication of non-existent entities, such as fake merchants or products, or the provision of attributes that conflict with real-world merchant information. Content Hallucination involves ungrounded or factually incorrect claims within a general context, such as misleading knowledge or suggestions that lack a factual basis.

The data construction followed a human-in-the-loop distillation pipeline. We initially conducted manual annotation on 2,000 samples to establish gold-standard hallucination patterns. These insights were then formalized into a set of structured evaluation prompts (as illustrated in Figure~\ref{fig:prompt_hallucination}), which served as the consistent labeling criteria for LLM-based distillation. This process resulted in a dataset of 9,391 real-world samples, where each entry is structured as a response paired with a corresponding list of identified hallucination issues. As summarized in Table~\ref{tab:hallucination_distribution}, the database contains 3,066 (33\%) instances of item hallucination and 1,188 (12\%) instances of content hallucination. Notably, 6,165 (65\%) of the samples are factually accurate responses. Since a single response may contain both types of hallucinations, the aggregate percentage surpasses 100\%. This composition allows the reward model to contrast incorrect outputs against verified positive examples, improving its ability to detect subtle factual errors in complex tasks.

\begin{table}[t]
    \centering
    \caption{Distribution of Hallucination Types in Factual Consistency Database.}
    \setlength{\aboverulesep}{0pt}
    \setlength{\belowrulesep}{0pt}
    \renewcommand{\arraystretch}{1.5}
    
    \begin{tabularx}{\linewidth}{C{3.5cm}||c|Y}
        \toprule
        \rowcolor{titlegray} \textbf{Hallucination Type} & \textbf{Percentage} & \textbf{Retrievals} \\ \midrule \midrule
        Item Hallucination & 33\% & 3,066 \\ \hline
        \rowcolor{tablegray} Content Hallucination & 12\% & 1,188 \\ \hline
        No Hallucination & 65\% & 6,165 \\ \bottomrule
    \end{tabularx}
    \label{tab:hallucination_distribution}
\end{table}

\subsection{Query-Item Relevance Database Construction}\label{sec:section_relev}
Beyond instruction adherence and factual consistency, the efficacy of an agentic recommender inherently depends on its ability to identify items that are semantically and functionally relevant to user queries. Without ensuring query-item relevance, the agent risks exploring an expansive space of irrelevant candidates, resulting in sparse and noisy feedback that hinders learning. To address this, we construct a specialized relevance database to evaluate and train agent proficiency in identifying relevant items.

\subsubsection{Data Collection}
We sampled over 20,000 interactions across six service categories from Meituan: Dining (35\%), Lifestyle (22\%), Shopping
(20\%), Tourism (9\%), Accommodation (9\%), and Healthcare (5\%), therefore enhancing the general relevance knowledge for model training, especially in category-specific attributes. Following the verification criteria (detailed in Figure ~\ref{fig:prompt_relev}), we evaluated relevance across seven primary dimensions and categorized each query-item pair into three levels: fully relevant, weakly relevant, and irrelevant.

To ensure high-quality labeling at scale, we employed a human-in-the-loop distillation pipeline. A seed set of 2,000 instances was first manually annotated to align the labeling behavior of GPT-4.1, which served as the teacher model for automated distillation using prompts synchronized with manual standards. This process yielded a final dataset of 19,456 instances, each structured as a triplet containing the user query, the candidate item response, and the corresponding relevance score. The final score distribution is detailed in Table~\ref{tab:relev_score}.

\begin{table}[htbp]
    \centering
    \small
    \captionsetup{justification=centering, skip=10pt}
    \caption{Evaluation Frameworks for Relevance Assessment}
    \label{tab:relev_score}

    \setlength{\aboverulesep}{0pt}
    \setlength{\belowrulesep}{0pt}
    \renewcommand{\arraystretch}{1}

    {\centering \textbf{(a) Evaluation Dimensions} \par}
    
    \begin{tabularx}{\linewidth}{C{2.8cm}||Y}
        \toprule
        \rowcolor{titlegray} \textbf{Evaluation Dimension} & \textbf{Dimension Description} \\ \midrule \midrule
        Category & Core business or product type (e.g., Dining vs. Hotel; Chinese vs. Western cuisine). \\ \hline
        \rowcolor{tablegray} Entity & Specific identifiers for merchants or brands as explicitly required in the user query. \\ \hline
        Geospatial Context & Precise location requirements, including specific landmarks, administrative districts, or relative distance constraints. \\ \hline
        \rowcolor{tablegray} Rating & Quantitative requirements regarding merchant or product reputation (e.g., "High rating," "Above 4.0"). \\ \hline
        Price & Monetary requirements, including price ranges, total budgets, or per-capita expenditure. \\ \hline
        \rowcolor{tablegray} Available Time & Temporal constraints such as business hours, real-time availability, or estimated delivery speed. \\ \hline
        User Persona & Scenario-specific requirements (e.g., student discounts, kid-friendly venues, or pet-friendly environments). \\ \bottomrule
    \end{tabularx}

    \vspace{10pt}
    
    {\centering \textbf{(b) Evaluation Score Distribution} \par}
    
    \begin{tabularx}{\linewidth}{C{2.5cm}||c|Y}
        \toprule
        \rowcolor{titlegray} \textbf{Evaluation Score} & \textbf{Percentage} & \textbf{Evaluation Criteria} \\ \midrule \midrule
        0 (Irrelevant) & 22\% & \textbf{Category Mismatch}: Category inconsistency or failure to align with the query's primary intent. \\ \hline
        \rowcolor{tablegray} 1 (Weakly Relevant) & 41\% & \textbf{Partial Alignment}: Category matches, but one or more core dimensions (Entity, Location, or Key Constraints) are unsatisfied. \\ \hline
        2 (Fully Relevant) & 37\% & \textbf{Full Satisfaction}: All explicit core requirements and implicit constraints (Numerical/Geospatial) are strictly met. \\ \bottomrule
    \end{tabularx}
\end{table}

\subsection{User Behavior Database Construction}

Accurately capturing user behavior is fundamental to the performance of agentic recommender systems. Traditional approaches often suffer from sparse and coarse-grained behavioral signals, as they typically rely on mapping latent user intents directly to final outcomes like clicks or orders. To address these limitations, we constructed a large-scale database that integrates comprehensive behavior prediction with a specialized item ranking sub-database. By enriching behavioral data with fine-grained user profiles and process-level ranking labels, this database provides a continuous feedback loop from intermediate preference to terminal actions, thereby mitigating positional bias and enhancing reward modeling efficiency.

\subsubsection{Data Collection}

We curated a large-scale dataset by sampling over 1 million real-world interactions, comprising 960,862 samples for behavior prediction and a refined subset of 75,648 samples for item ranking. Each entry is structured as a comprehensive tuple containing the user profile, user query, recommendation candidates, and the corresponding behavior prediction. To precisely characterize user intent, we deconstructed user profiles into four granular dimensions: basic demographics, consumer profile, long-term preferences, and real-time context (detailed in Figure~\ref{fig:prompt_behavior}).

\subsubsection{Interaction Density Analysis}

The necessity of incorporating item ranking is underscored by the inherent sparsity of terminal behavioral signals. As shown in Table~\ref{tab:interaction_density}, statistical analysis of our dataset reveals that out of 4,163,922 total item exposures, only 23.1\% resulted in positive behavioral outcomes (clicks or orders). This pronounced sparsity justifies the inclusion of process-level ranking labels, which offer dense, relative preference signals to complement the sparse terminal feedback. This multi-layered structure provides a robust resource for training reward models that are sensitive to both subtle preference variations and final conversion outcomes.

\begin{table}[t]
    \centering
    \caption{Interaction density and signal distribution in the User Behavior Database.}
    \setlength{\aboverulesep}{0pt}
    \setlength{\belowrulesep}{0pt}
    \renewcommand{\arraystretch}{1.5}
    
    \begin{tabularx}{\linewidth}{C{2cm}||c|Y}
        \toprule
        \rowcolor{titlegray} \textbf{Granularity} & \textbf{User Preference} & \textbf{Percentage} \\ \midrule \midrule
        \multirow{2}{*}{List-wise} & Interested (At least one click) & \textbf{25.0}\% \\ \cline{2-3}
                                            & Uninterested (Zero click)       & 75.0\% \\ \hline
        \multirow{2}{*}{Item-wise} & Interested (Clicked/Ordered)    & \textbf{23.1}\% \\ \cline{2-3}
                                            & Uninterested (Exposed only)     & 76.9\% \\ \bottomrule
    \end{tabularx}
    \label{tab:interaction_density}
\end{table}


\section{RecRM-RL}\label{sec:section_RL}

\begin{figure*}
    \centering
    \includegraphics[width=\linewidth]{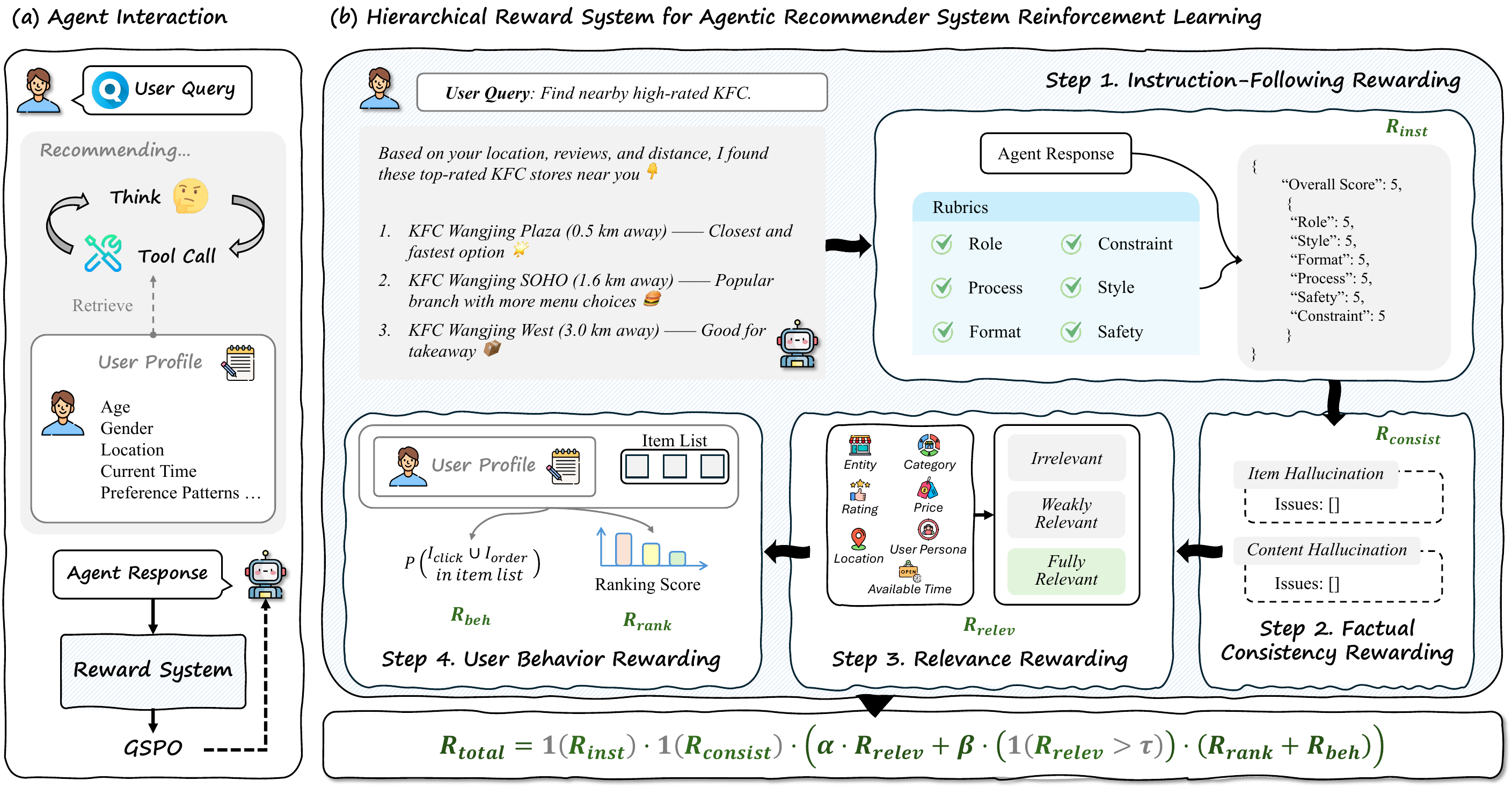}
    \caption{Overview of proposed RecRM-RL framework.}
    \label{fig:method}
\end{figure*}

Building on RecRM-Bench, we develop a systematic reinforcement learning framework for agentic recommender systems (shown in Figure ~\ref{fig:method}). This framework adopts the ReAct paradigm \cite{yao2023react} to synergize reasoning and acting for complex recommendation tasks. While the ReAct architecture involves iterative steps of thought and action, our optimization strategy focuses on the quality of the final response rather than internal tool-calling trajectories, as the final output directly determines the overall performance of the recommendation system. By leveraging the four previously constructed databases to train independent reward models, we define the essential dimensions for agent evaluation, including instruction following, retrieval relevance, factual consistency, and user behavior prediction. These individual components are then integrated into a hybrid multi-objective reward system, providing the agent with comprehensive feedback to synergistically enhance agent performance. Detailed descriptions of each reward model and the overall training objective are provided in the following sections.

\subsection{Instruction Following Reward Model}

To ensure that the final recommendation responses adhere to the required formats and constraints, we develop an instruction-following reward model by performing supervised fine-tuning (SFT) on a Qwen3-8B backbone. This model leverages the specialized database introduced in \S ~\ref{sec:section_if}, which contains fine-grained rubrics for diverse instruction scenarios. Specifically, we concatenate these rubrics with a dedicated judge system prompt to evaluate the compliance degree of the agent response. This evaluation covers both overall performance and individual constraint dimensions. Formally, for a given judge prompt $p_{if}$ (containing the rubrics) and collected online response $r_i$, the model outputs the verification reason $t_i$ and the final judge score $s_i$ (including overall score and dimensional-specific scores):
\begin{equation}
(t_i, s_i)=f_{if}([p_{if},r_i];\Phi)
\end{equation}
where $\Phi$ represents the training parameters of LLM.

By optimizing the cross-entropy loss during the fine-tuning process, the reward model learns to internalize complex requirements, thereby eliminating the need for manual or task-specific rule checking. The finetuned model serves as an automated evaluator that reflects how well the agent follows the prompt instructions.

\subsection{Factual Consistency Reward Model}
Ensuring factual consistency and mitigating hallucinations are fundamental prerequisites for the reliability of agentic recommendation systems. To provide robust grounding for the agent's outputs, we develop a specialized factual consistency reward model by training a Qwen3-8B backbone on our factual consistency database. This model is designed to act as a critical verifier, identifying fine-grained discrepancies between the agent's generated content and the tool output or general knowledge. Formally, given a judge prompt $p_{consist}$ defining the principles for hallucination detection and the agent response $r_i$, the model identifies the set of factual inconsistencies $\mathcal{Q}$:
\begin{equation}
    \mathcal{Q} = f_{consist}(p_{consist}, r_i;\Phi)
\end{equation}
where $\Phi$ represents the model parameters optimized via a sequence-level cross-entropy objective. This optimization maximizes the model's sensitivity to subtle hallucinations, thereby providing a high-fidelity reliability signal that guides the agent toward more faithful reasoning trajectories.

\subsection{Query-Item Relevance Reward Model}
To evaluate the alignment between recommended items and user queries, we develop a specialized relevance reward model by fine-tuning Qwen3-14B via SFT. This model utilizing the comprehensive collection of human-annotated and LLM-distillation data (as introduced in \S ~\ref{sec:section_relev}) is trained to generate a detailed justification followed by a specific relevance score. Such training enables the model to perform a deep semantic analysis of the relationship between item attributes and specific user requirements. This reward model also serves to narrow the action space during reinforcement learning. By penalizing irrelevant items, it prevents inefficient exploration within the irrelevant sample space and ensures that the agent remains grounded in user intent. Formally, given a judge prompt $p_{relev}$ outlining the relevance principles, a user query $q_i$, and the agent response $r_i$, the model outputs the verification reason $t_i$ and a final judge score $s_i$:
\begin{equation}
(t_i, s_i)=f_{relev}([p_{relev}, q_i, r_i];\Phi)
\end{equation}
where $\Phi$ represents the training parameters of LLM and the model is also optimized via cross-entropy loss to maximize the likelihood of expert-level reasoning and scoring trajectories.

\subsection{User Behavior Reward Model}
The effectiveness of the reinforcement learning process depends on a reward model capable of accurately predicting user decisions by integrating user profiles with historical interaction patterns. To address the inherent sparsity of direct user feedback, we develop a two-stage reward mechanism that provides dense supervision signals to stabilize agent training.

\subsubsection{Item Ranking}
To evaluate the initial candidates retrieved by tool responses, we implement a comprehensive ranking-based reward that provides intermediate feedback. For this task, we extract 82,993 high-quality samples to fine-tune a Qwen3-Reranker-0.6B model. Formally, given the user profile $u_i$, query $q_i$, and a candidate set $\mathcal{C}_i = \{c_{i,1}, c_{i,2}, \dots, c_{i,K}\}$, the reranker $f_{R}$ computes a relevance score $s_{i,j}$ for each candidate $c_{i,j}$:

\begin{equation}
s_{i,j} = f_{rank}(p_{rank}, u_i, q_i, c_{i,j}; \Phi)
\end{equation}
where $p_{rank}$ denotes the ranking prompt, and $\Phi$ represents the model parameters. The input format and prompt structure of $p_{rank}$ are kept strictly consistent with the reranker's original prompt, as illustrated in Figure ~\ref{fig:prompt_item_ranking}.

The optimization objective integrates both item-wise click-through prediction and list-wise ranking signals via a dual-loss mechanism:
\begin{equation}
\mathcal{L} = \mathcal{L}_{cls}(s_{i,j}, y_{i,j}) + \lambda \mathcal{L}_{rank}(\mathbf{s}_i, y_{i,pos})
\end{equation}
Specifically, $\mathcal{L}_{cls}$ is a binary cross-entropy loss for item-wise click prediction, while $\mathcal{L}_{rank}$ is a multi-class cross-entropy loss that identifies the ground-truth positive item $y_{i,pos}$ from the candidate set $\mathcal{C}_i$. This dual-objective mechanism enables the reward model to capture both the absolute probability of a click and the relative ranking relationships, ensuring that the agent prioritizes high-potential candidates during the reasoning process.

\subsubsection{Behavior Prediction}
To capture the core objective of personalized recommendation, we fine-tune a Qwen3-0.6B model on the full dataset of 1,000,000 behavioral feedback sequences. The primary challenge here is the extremely low interaction density, which makes individual item-level signals highly susceptible to noise. To mitigate this, we employ a specialized list-based evaluation logic. This approach shifts the focus from brittle point-wise predictions to list-level alignment, effectively capturing whether the agent response encompasses the user's potential interests within the candidate space. Formally, given the user profile $u_i$ and query $q_i$, the model $f_{beh}$ predicts the likelihood of the ground-truth positive item $c_{i,pos}$ being included in the generated recommendation list $\mathcal{C}_i$:

\begin{equation}
\hat{y}_i = f_{beh}([p_{beh}, u_i, q_i, \mathcal{C}_i]; \Phi)
\end{equation}
where $p_{beh}$ is the feedback evaluation prompt and $\Phi$ represents the model parameters. The input format and prompt structure for $p_{beh}$ are designed to be consistent with the original inference prompt, as detailed in Figure ~\ref{fig:prompt_behavior}.

Under this scheme, a prediction is considered successful if $c_{i,pos} \in \mathcal{C}_i$. The model is optimized via cross-entropy loss to maximize the alignment between predicted probabilities and actual user outcomes. By optimizing this list-based objective, the reward model captures essential user-item interactions and latent preferences, providing a precise and explicit reward signal that aligns the final agent output with historical user patterns.

\subsection{RecRM-RL Training Objective}\label{sec:section_RL_final}
We explore the development of an agentic recommendation system by fine-tuning Qwen3-235B-A22B, providing a reference methodology for integrating multi-dimensional reward models under GSPO. To effectively integrate the multi-dimensional reward models introduced previously, we propose a hierarchical reinforcement learning objective. This structure implements a strict sequential gating mechanism, ensuring that the agent prioritizes foundational reliability before pursuing complex user behavior optimization. The training process follows a three-stage validation pipeline. In the first stage, the response $r_i$ undergoes a compliance check via the instruction following reward model; any violation of the required format sets the indicator $C_{\text{inst}}$ to 0. In the second stage, the factual consistency reward model verifies the presence of hallucinations or discrepancies relative to the item attributes. We define a binary constraint $C_{\text{consist}}$ that equals 1 only if the response is entirely consistent with the provided tool outputs and free of hallucinations, and 0 otherwise. 

Only after passing these foundational gates (i.e., $C_{\text{inst}} \cdot C_{\text{consist}} = 1$) does the system proceed to evaluate the semantic and behavioral quality of the response. First, the relevance reward model calculates $R_{\text{relev}}$ to ensure the recommended items align with the user's query. If this relevance score $R_{\text{relev}}$ exceeds a predefined threshold $\tau$, the system further activates the behavioral rewards. These include the item ranking reward $R_{\text{rank}}$, which evaluates the relative preference among candidates, and the behavior prediction reward $R_{\text{beh}}$, which measures the likelihood of a final user interaction. This hierarchical mechanism ensures that the agent only receives behavioral optimization signals when the response is factually sound and contextually relevant, effectively preventing the model from converging toward hallucinated or irrelevant regions of the item space.

To ensure training stability, each reward component is normalized to the range $[0, 1]$. The overall composite reward $R_{\text{total}}$ is formally defined as:
\begin{equation}
R_{\text{total}} = C_{\text{inst}} \cdot C_{\text{consist}} \cdot \Big( \alpha \cdot R_{\text{relev}} + \beta \cdot \mathbbm{1}\left(R_{\text{relev}} > \tau\right) \cdot (R_{\text{rank}} + R_{\text{beh}}) \Big)
\end{equation}
where $\alpha$ and $\beta$ are weighting coefficients that determine the relative importance. This hierarchical mechanism forces the agent to prioritize fundamental alignment, effectively pruning the search space and preventing the optimization process from converging toward irrelevant regions of the item space.

\section{Evaluation}
\subsection{Evaluation of RecRM-Bench}
\subsubsection{Evaluated Models}
We evaluate state-of-the-art models, including both thinking and non-thinking models, across each database of RecRM-Bench. The evaluated models encompass GPT 4.1~\cite{gpt-4.1}, the LongCat Series (LongCat Flash Cat and LongCat Flash Thinking)~\cite{team2025longcat}, DeepSeek V3.2~\cite{liu2025deepseek}, and Qwen3-Max~\cite{qwen3-max}. Furthermore, we evaluate our optimized reward models (denoted as Ours), described in \S~\ref{sec:section_RL}, to provide a direct performance comparison against the zero-shot baselines. Following the verification prompts, we parse the model responses to extract scores for metric calculation.

For the tasks of instruction following, factual consistency, and query item relevance, where the expected outputs are numerical, we evaluate the models using Accuracy (ACC) and F1-score. For the item ranking and behavior prediction subsets, we utilize Acc and Area Under the Curve (AUC) as the primary metrics. Additionally, Hit Rate (HR) at various depths is introduced to evaluate the ranking performance, following ~\cite{shang2025agentrecbench}. To ensure a fair and comprehensive assessment, we report the average of HR@1, HR@3, and HR@5 as the final ranking metric.

\subsubsection{Data Quality}\label{sec:section_data_quality} The reliability of our constructed databases is critical to downstream 
performance. We therefore perform dedicated validation studies on the 
Instruction Following database and the Query-Item Relevance database, focusing on rubric effectiveness, LLM distillation faithfulness, and per-class relevance prediction consistency.

\paragraph{Instruction Following Database Validation.} To construct the Instruction Following database, we design scoring rubrics and synthesize data for augmentation. We first verify the reliability of these rubrics by evaluating the score margin between perfect and synthesized imperfect responses. To ensure fairness and robustness, we selected three top-performing evaluators from our benchmark: GPT-4.1, Longcat-Flash-Chat, and Longcat-Flash-Thinking. As shown in Table ~\ref{tab:if_model_score_comparison}, the scores for synthesized responses are significantly lower than those for perfect ones, demonstrating that our rubrics effectively enable the model to distinguish response quality. To further assess the reliability of the synthesized data, we quantify human-machine agreement using the weighted Cohen’s $\kappa$. Results in Table \ref{tab:if_human_machine_agreement} show that the overall instruction following score reaches $\kappa>0.61$, indicating high data reliability. Regarding dimensional consistency, constraint compliance and process compliance achieve the highest agreement ($\kappa>0.87$). In contrast, content quality compliance and style compliance yield relatively lower scores. This aligns with the observation that models excel at following explicit processes and constraints but occasionally struggle with generating high-quality content or adhering to specific styles.

\begin{table*}[t] 
    \centering
    \caption{Performance Comparison of Different Models across Four Tasks.
    The top and worst performing results are highlighted in 
    \colorbox{topcolor}{\textbf{red}} (1\textsuperscript{st}) and 
    \colorbox{botcolor}{\textbf{blue}} (bottom) backgrounds, respectively.}
    \label{tab:model_performance}
    \scriptsize 
    \renewcommand{\arraystretch}{1.2}
    
    \begin{tabular*}{\textwidth}{@{\extracolsep{\fill}}l|cc|cc|cc|ccc|cc}

        \toprule[1.5pt]
        
        \multirow{2}{*}{\textbf{Models}}
        & \multicolumn{2}{c|}{\textbf{Instruction Following}} 
        & \multicolumn{2}{c|}{\textbf{Factual Consistency}} 
        & \multicolumn{2}{c|}{\textbf{Query-Item Relevance}} 
        & \multicolumn{3}{c|}{\textbf{Item Ranking}} 
        & \multicolumn{2}{c}{\textbf{Behavior Prediction}} \\ 

        \cline{2-12}

        & \textbf{ACC (\%)} & \textbf{F1-Score (\%)} 
        & \textbf{ACC (\%)} & \textbf{F1-Score (\%)} 
        & \textbf{ACC (\%)} & \textbf{F1-Score (\%)} 
        & \textbf{ACC (\%)} & \textbf{AUC (\%)} & \textbf{HR (\%)} 
        & \textbf{ACC (\%)} & \textbf{AUC (\%)} \\ 
        
        \midrule
        
        \textbf{GPT-4.1} 
            & 55.47 & 58.33 
            & 62.96 & 77.27
            & 79.26 & 79.48 
            & 42.54 & 51.32 & 75.37 
            & 39.28 & 49.31 \\ \hline
        
        \textbf{LongCat-Flash-Chat} 
            & 64.84 & 65.19
            & 67.34 & 80.48
            & \botval{73.18} & \botval{72.82}
            & 62.48 & 54.93 & 82.93 
            & 46.49 & 52.22 \\
        
        \textbf{LongCat-Flash-Thinking} 
            & 43.75 & 48.16 
            & 64.98 & 78.78 
            & 75.97 & 76.17 
            & 34.09 & 57.89 & \botval{70.00}
            & \botval{31.02} & \botval{45.41} \\ \hline
        
        \textbf{Deepseek-V3.2} (w/o thinking) 
            & 30.47 & 29.94
            & 43.43 & 60.56
            & 74.60 & 74.76 
            & 52.32 & 50.57 & 82.53 
            & 47.80 & 52.10 \\
        
        \textbf{Deepseek-V3.2} (w/ thinking) 
            & 35.29 & 39.50 
            & \botval{41.41} & \botval{58.57}
            & 75.22 & 75.57 
            & \botval{29.55} & 57.04 & 72.42 
            & 35.17 & 50.76 \\ \hline
        
        \textbf{Qwen3-Max} (w/o thinking) 
            & 36.80 & 40.33 
            & 53.20 & 69.45
            & 76.64 & 77.02 
            & 50.11 & 50.18 & 77.17 
            & 42.79 & 52.08 \\

        \textbf{Qwen3-Max} (w/ thinking) 
            & \botval{26.67} & \botval{26.64}
            & 56.57 & 72.26
            & 75.89 & 76.16
            & 50.16 & \botval{50.01} & 78.42
            & 44.47 & 48.50 \\ \hline
        
        \textbf{Ours} 
            & \topval{72.66} & \topval{72.40} 
            & \topval{70.71} & \topval{82.84}
            & \topval{89.36} & \topval{89.12}
            & \topval{86.78} & \topval{86.32} & \topval{83.67}
            & \topval{77.78} & \topval{81.46} \\
        
        \bottomrule[1.5pt]
    \end{tabular*}
\end{table*}

\paragraph{Query-Item Relevance Database Validation.} To construct the Query-Item Relevance database, we implemented a two-stage pipeline involving expert annotation followed by LLM distillation. To validate the reliability of this distillation process, we conducted a human-machine alignment study on a held-out sample. The results yield a weighted $\kappa=0.71, \rho=0.76$, with a 77\% raw agreement, indicating substantial alignment between the distilled model and human experts. Beyond global correlation, we analyzed the per-class F1-scores to mitigate the risk of polarity conflicts (e.g., confusing irrelevant with fully relevant). The model achieves an F1-score of 79.45\% for irrelevant and 81.36\% for fully relevant items, compared to 70.59\% for the weakly relevant. This performance distribution demonstrates that inconsistencies are primarily confined to boundary cases where semantic ambiguity is inherently higher, whereas the model maintains robust, high-confidence predictions in unambiguous scenarios.

\begin{table}[t]
    \centering
    \caption{Human-Machine Agreement: Cohen's $\kappa$ and Spearman's $\rho$ on synthesized data in Instruction Following database.}
    \label{tab:if_human_machine_agreement}
    \small
    \renewcommand{\arraystretch}{1.2}
    \begin{tabular}{l|c|c}
        \toprule[1.5pt]
        \textbf{Dimension} & \textbf{Cohen's $\kappa$} & \textbf{Spearman's $\rho$} \\
        \midrule
        Overall               & 0.6574 & 0.7081 \\
        Role Compliance       & 0.8254 & 0.8339 \\
        Process Compliance    & 0.8771 & 0.9344 \\
        Format Compliance     & 0.8225 & 0.8049 \\
        Content Quality       & 0.7754 & 0.7771 \\
        Constraint Compliance & 0.8856 & 0.7149 \\
        Style Compliance      & 0.7700 & 0.7239 \\
        \midrule
        \rowcolor{lightblue} \textbf{Average}      & \textbf{0.8019} & \textbf{0.7853} \\
        \bottomrule[1.5pt]
    \end{tabular}
\end{table}

\begin{table}[t]
    \centering
    \caption{Score Discriminability: Average instruction following scores on Collected Perfect Response vs. Synthesized Imperfect Response in Instruction Following database.}
    \label{tab:if_model_score_comparison}
    \footnotesize
    \renewcommand{\arraystretch}{1.2}
    \begin{tabular}{l|c|c}
        \toprule[1.5pt]
        \textbf{Model} & \textbf{\makecell{Synthesized Imperfect \\ Response}} & \textbf{\makecell{Collected Perfect \\ Response}} \\
        \midrule
        GPT-4.1                & 2.19 & 4.53 \\
        LongCat-Flash-Chat     & 2.93 & 4.92 \\
        LongCat-Flash-Thinking & 2.22 & 3.89 \\
        \midrule
        \rowcolor{lightblue} \textbf{Average} & \textbf{2.45} & \textbf{4.45} \\
        \bottomrule[1.5pt]
    \end{tabular}
\end{table}

\subsubsection{Evaluation Performance}\label{sec:section_comparison}

As shown in Table ~\ref{tab:model_performance}, GPT-4.1 achieves the leading query-item relevance accuracy of 79.26\%, whereas LongCat-Flash-Chat dominates the instruction following (64.84\%), item ranking (62.48\% accuracy, 82.93\% HR), and factual consistency (67.34\%) tasks. Concurrently, DeepSeek-V3.2 exhibits the highest proficiency in user behavior prediction with an accuracy of 47.80\%. The validation results also reveal a distinct performance trade-off that thinking models demonstrate superior performance in complex query-item relevance tasks, while non-thinking models exhibit superior performance in the remaining databases.

Despite these individual strengths, the overall performance of these models remains limited and imbalanced. The zero-shot results are significantly lower than those of models optimized via SFT, particularly in item ranking and behavior prediction metrics. Specifically, our trained model (Ours) achieves a substantial performance leap, reaching 89.36\% accuracy in query-item relevance and 86.78\% in item ranking, significantly outperforming the zero-shot baselines. Beyond the gap between zero-shot and SFT performance, results across different databases vary substantially. While these models can achieve accuracies exceeding 70$\%$ in query-item relevance, their performance drops sharply in other databases. This pronounced imbalance further emphasizes the importance of implementing a hybrid reward function to develop comprehensive and highly capable recommendation agents. This disparity further highlights the inherent limitations of general LLMs on RecRM-Bench and underscores the necessity of specialized reward model training. 

\subsection{Performance of RecRM-RL}
\begin{figure}
    \centering
    \includegraphics[width=\linewidth]{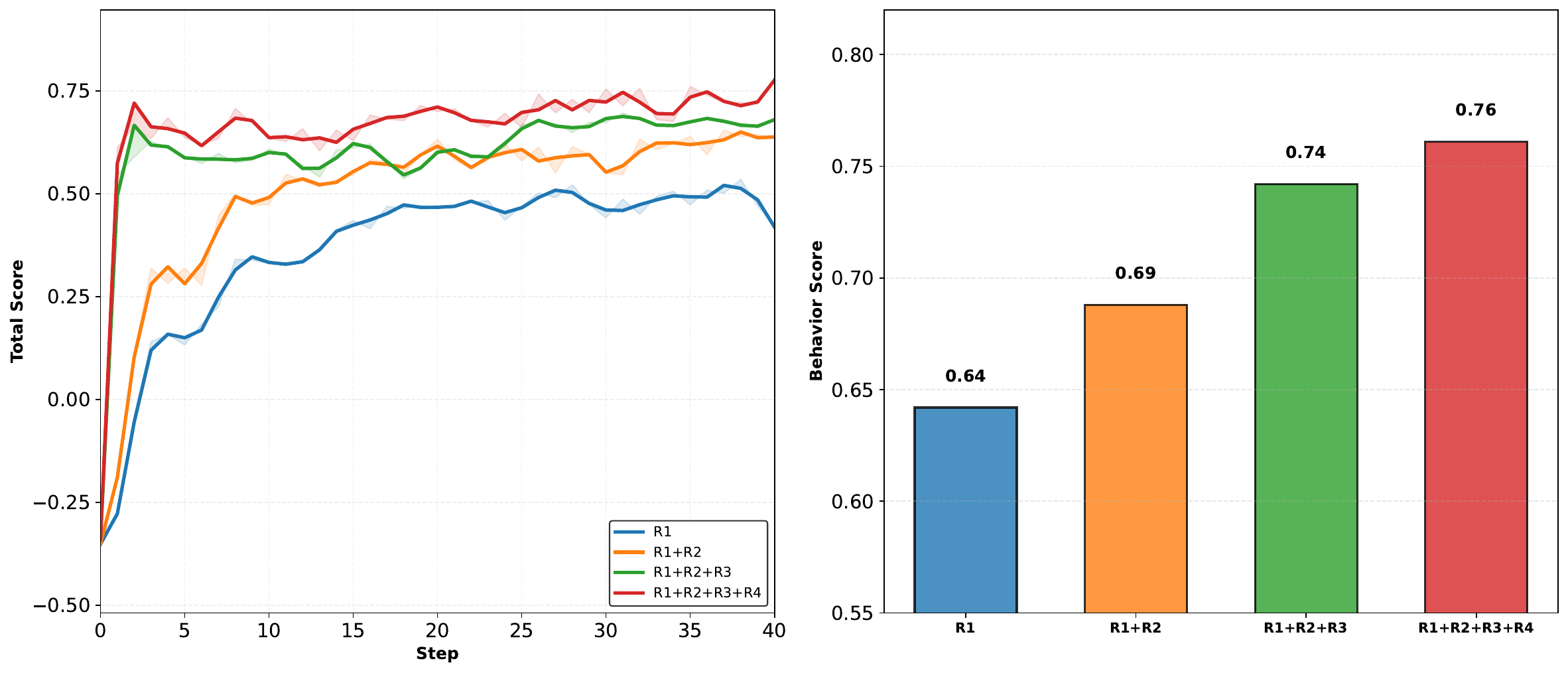}
    \caption{Overall performance of proposed RecRM-RL framework. (a) The training process of RecRM-RL, where the total score represents the final reward score; (b) The final user behavior prediction accuracy of models training on different strategies.}
    \label{fig:rm_results}
\end{figure}

Beyond evaluating the performance of a single reward model trained on RecRM-Bench, we further assess the proposed hierarchical RL framework introduced in \S ~\ref{sec:section_RL_final}. As illustrated in Figure ~\ref{fig:rm_results}(a), where R1 represents the baseline, R2 incorporates Instruction Following, R3 adds Query-item Relevance, R4 includes User Behavior, and R5 introduces Factual Consistency, the integration of these structured signals substantially accelerates the convergence of the training process compared to the baseline. Notably, the most significant gain in learning efficiency occurs after integrating R3 (Query-Item Relevance), which demonstrates that recognizing user intent is a prerequisite for effective exploration. This trend suggests that intermediate rewards effectively mitigate the reward sparsity challenge by providing explicit optimization paths, thereby minimizing inefficient exploration during the early phases of reinforcement learning.

Regarding the impact on final behavioral predictions (shown in Figure ~\ref{fig:rm_results}(b)), the final behavior score have improved progressively by 19\% as all rewards included. This enhancement is particularly evident after integrating Relevance reward (R3) by 7.8\%. While the final behavior prediction reward provides relatively small gains, it further illustrates that the system constructs a more reliable internal representation of the user state by accurately identifying item relevance, which is a prerequisite for precise behavioral forecasting. Furthermore, while intermediate rewards like Instruction Following (R2) and Factual Consistency (R5) lack a direct functional dependency on the final click, they serve as essential search-space optimizers. By stabilizing the early stages of training and pruning ineffective or illogical paths, these rewards effectively reduce optimization noise. This allows the model to more efficiently identify the optimal recommendation policy within a valid, logical candidate pool, ensuring the system is not only predictive of user actions but also instruction-compliant and factually robust.

\subsection{Ablation Study}
\subsubsection{Impact of Data Augmentation}
For the Instruction Following database, we propose to synthesize dimension-specific data for data augmentation. To evaluate the effectiveness of these synthesized data, we evaluate the data quality mainly from the performance of reward model, since it directly affects further agentic recommender training. As shown in Table ~\ref{tab:if_syn_ablation}, the inclusion of synthesized data leads to a 15.63\% improvement in the prediction accuracy for overall score. This result demonstrates that the synthetic samples effectively supplement the original data distribution and enhance the model's performance. Specifically, while substantial gains are observed in format and role compliance, the improvements in content quality (0.78\%) and style (a 0.78\% reduction) are relatively small. These findings are consistent with observations in human-machine alignment (see \S ~\ref{sec:section_data_quality}), suggesting that improving these nuanced and subjective dimensions requires more sophisticated synthesis strategies.

\begin{table}[t]
    \centering
    \caption{Impact of synthetic data augmentation on Instruction Following accuracy across dimensions.}
    \label{tab:if_syn_ablation}
    \small
    \renewcommand{\arraystretch}{1.2}
    \begin{tabular}{l|c|c|c}
        \toprule[1.5pt]
        \textbf{Dimension} & \textbf{wo/ syn ACC. (\%)} & \textbf{w/ syn ACC. (\%)} & \textbf{$\Delta$ (\%)} \\
        \midrule
        Role            & 75.00 & 83.59 & \textcolor{green!60!black}{\textbf{+8.59}}  \\
        Process         & 80.47 & 84.38 & \textcolor{green!60!black}{\textbf{+3.91}}  \\
        Format          & 60.94 & 71.88 & \textcolor{green!60!black}{\textbf{+10.94}} \\
        Content Quality & 66.41 & 67.19 & \textcolor{green!60!black}{\textbf{+0.78}}  \\
        Constraint      & 83.59 & 88.28 & \textcolor{green!60!black}{\textbf{+4.69}}  \\
        Style           & 75.00 & 74.22 & \textcolor{red!70!black}{\textbf{-0.78}}   \\
        \rowcolor{lightblue} \textbf{Overall Score} & 57.03 & 72.66 & \textcolor{green!60!black}{\textbf{+15.63}} \\
        \bottomrule[1.5pt]
    \end{tabular}
\end{table}

\subsubsection{Impact of Base Retrievers}
\begin{figure}
    \centering
    \includegraphics[width=\linewidth]{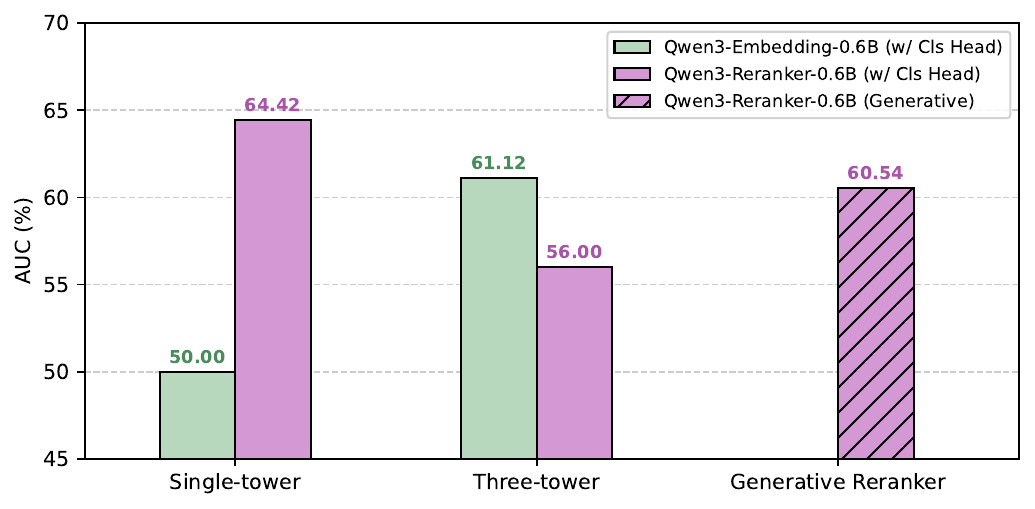}
    \caption{Ablation study on reranker backbone architectures.}
    \label{fig:reranker_backbone}
\end{figure}

The item ranking reward is important in the user behavior reward model. To identify the optimal architecture, we conduct a comprehensive analysis of various backbones. Our evaluation compares two primary backbones, Qwen3-Embedding-0.6B and Qwen3-Reranker-0.6B, across classification-based heads (single-tower and three-tower) and generative rerankers. As illustrated in Figure ~\ref{fig:reranker_backbone}, the single-tower Reranker achieves a 14\% AUC gain over the Embedding backbone due to its exhaustive cross-attention mechanism. However, this trend reverses in multi-tower configurations, as embedding models are better pre-trained for the disentangled representations required by late-fusion bottlenecks.

Despite the single-tower classification head achieving the best result, we ultimately adopt the generative reranker for our final framework. This selection prioritizes the instruction-based flexibility of generative models, enabling the system to incorporate multi-dimensional evaluation criteria through prompt engineering without structural re-design. Furthermore, this selection maintains architectural consistency with the primary generative agent, facilitating a unified semantic space for stable policy optimization.

\subsubsection{Impact of Model Size}

We investigate the impact of model scale on reward model performance by evaluating backbones of varying sizes, including Qwen3-0.6B, 8B, and 14B. As reported in Table \ref{tab:model_size_performance}, the 14B variant excels in the knowledge-intensive Query-Item Relevance task, where larger parameter counts facilitate the deep semantic understanding necessary for nuanced matching. In contrast, the 8B model emerges as the optimal scale for Instruction Following and Behavior Prediction. We attribute this to the high sparsity and noise inherent in behavioral data; while the 14B model may overfit idiosyncratic user noise, the 8B model offers a superior inductive bias by capturing generalized preference patterns. Similarly, for instruction following, the 8B scale provides sufficient reasoning depth to parse complex constraints while maintaining the efficiency required for specialized formatting tasks.

\begin{table}[t]
    \centering
    \caption{Performance scaling across model sizes. The top performing results are highlighted in \colorbox{topcolor}{\textbf{red}} (1\textsuperscript{st}) backgrounds.}
    \label{tab:model_size_performance}
    \scriptsize
    \begin{tabular}{l|cc|cc|cc}

        \toprule[1.5pt]

        \multirow{2}{*}{\textbf{Model Size}}
        & \multicolumn{2}{c|}{\textbf{Instruction Following}}
        & \multicolumn{2}{c|}{\textbf{Query-Item Relevance}}
        & \multicolumn{2}{c}{\textbf{Behavior Prediction}} \\

        \cline{2-7}

        & \textbf{ACC (\%)} & \textbf{F1 (\%)}
        & \textbf{ACC (\%)} & \textbf{F1 (\%)}
        & \textbf{ACC (\%)} & \textbf{F1 (\%)} \\

        \midrule

        \textbf{Qwen3-0.6B}
            & 69.53 & 70.34
            & 73.21 & 62.52
            & 70.17 & 72.18 \\ \hline

        \textbf{Qwen3-8B}
            & \topval{72.66} & \topval{72.40}
            & 83.79 & 79.93
            & \topval{70.30} & \topval{72.25} \\ \hline

        \textbf{Qwen3-14B}
            & 71.09 & 71.21
            & \topval{84.72} & \topval{81.22}
            & 42.63 & 41.02 \\

        \bottomrule[1.5pt]

    \end{tabular}
\end{table}

\subsection{Failure Case Analysis}
We investigate the primary factors limiting the performance of current state-of-the-art models on RecRM-Bench, with a specific focus on the Factual Consistency and Query-Item Relevance databases.

\begin{figure}
    \centering
    \includegraphics[width=\linewidth]{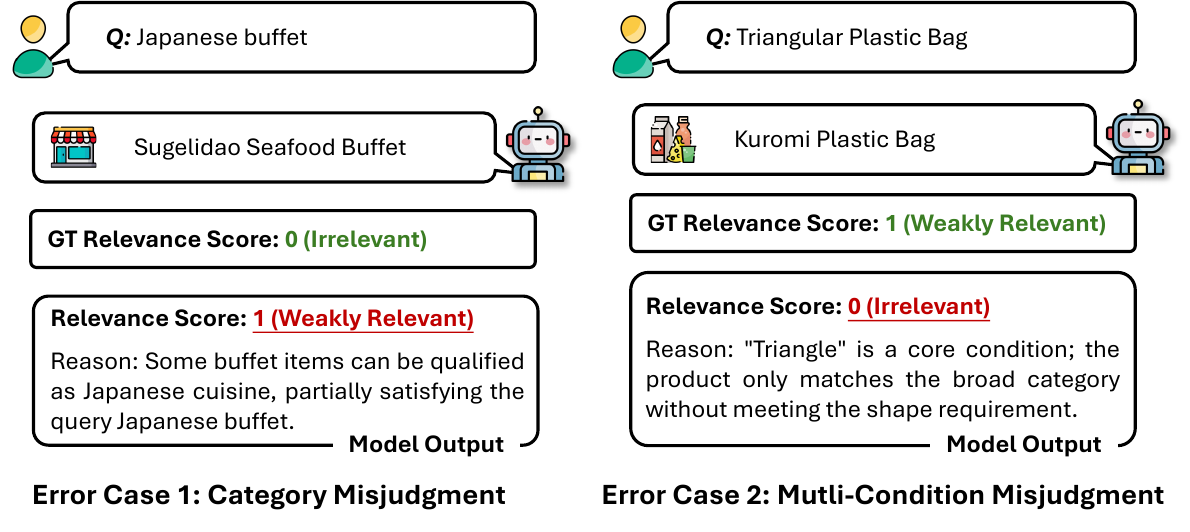}
    \caption{Representative failure cases in the Query-Item Relevance database.}
    \label{fig:failure_case_relev}
\end{figure}

\begin{figure}
    \centering
    \includegraphics[width=0.85\linewidth]{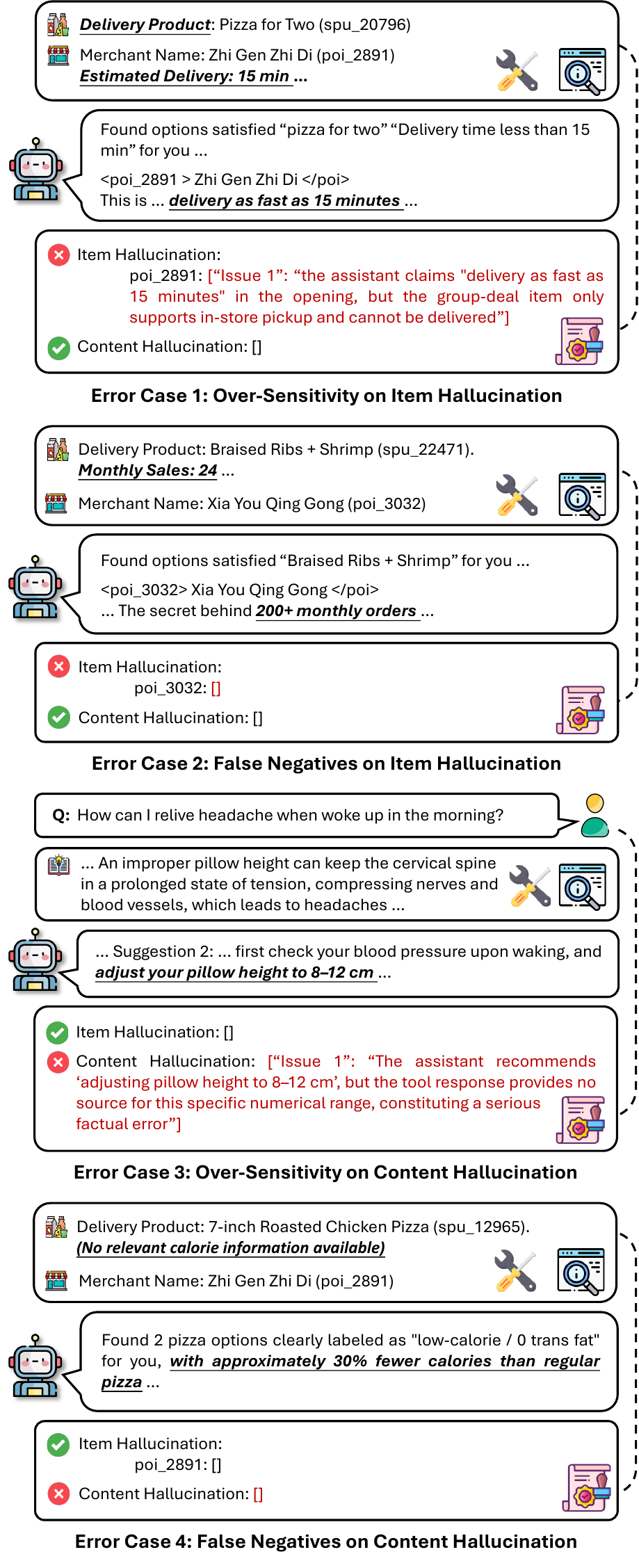}
    \caption{Representative failure cases in the Factual Consistency database.}
    \label{fig:failure_case_hallucination}
\end{figure}

The failure cases in the Factual Consistency database can be classified into four types: over-sensitivity on item hallucination (56.6\%), false negatives on item hallucination (20.2\%), over-sensitivity on content hallucination (16.1\%), and false negatives on content hallucination (7.1\%). Figure~\ref{fig:failure_case_hallucination} provides examples for each type. These results highlight a systematic failure mode: models frequently fail to perform rigorous cross-verification against the provided reference item attributes. This leads to either an over-reliance on subjective phrasing without grounding in the reference data or the generation of unwarranted inferences regarding item details. Consequently, strengthening the model’s capability to validate generated responses strictly against reference metadata is essential for enhancing reliability in agentic recommendation.

Failure cases in the Query-Item Relevance database primarily stem from two issues: multi-conditional intent misjudgement (34.4\%) and category misjudgement (25\%), where the former involves multiple conditions in user intent. Figure~\ref{fig:failure_case_relev} provides examples for each type. These failures reveal two critical model-level bottlenecks. The first is Category Misjudgment and Knowledge Gaps that models often struggle to distinguish primary from peripheral categories. For example, the model misinterprets a general seafood buffet as weakly relevant to a "Japanese buffet" query, reflecting insufficient domain-specific knowledge for accurate item-to-intent matching. The second bottleneck is Logical Inconsistency in Multi-condition Parsing that models oscillate between overly strict penalization of missing details and imprecise leniency, lacking a standardized framework for condition weighting. This instability reveals that enhancing the model's capacity for rigorous multi-constraint reasoning is indispensable for achieving robust alignment between complex user preferences and the final recommended items, ensuring the reliability of agentic recommendations.

\section{Conclusion}
In this paper, we propose \textsc{RecRM-Bench}, the first large-scale, comprehensive benchmark specifically designed for training multi-dimensional reward models in agentic recommender systems. Comprising 1,073,779 high-quality samples across four distinct sub-databases, RecRM-Bench provides explicit and granular guidance spanning \textit{Instruction Following}, \textit{Factual Consistency}, \textit{Query-Item Relevance}, and \textit{User Behavior}. Furthermore, we introduce RecRM-RL, a hierarchical reinforcement learning framework that demonstrates how these multi-dimensional reward signals can be effectively integrated to optimize agentic behavior. By establishing this foundational benchmark, we aim to bridge the gap between generative reasoning and personalized recommendation. We open-source RecRM-Bench to contribute to the development of next-generation recommender agents that are both factually reliable and deeply personalized.

\bibliographystyle{ACM-Reference-Format}
\bibliography{ref}

\appendix
\clearpage
\clearpage
\section{System Prompt Design for Data Generation and Reward Modeling}
This section shows the detailed prompt framework for dataset construction.

\begin{figure}[H]
    \centering

    \begin{tcolorbox}[
    colback=gray!5!white,
    colbacktitle=gray!30!white,
    colframe=black,
    boxrule=1pt,
    left=1mm, right=1mm,
    halign title=center,
    rounded corners,
    coltitle=black,
    title=Prompt for Instruction Following Data Augmentation, width=\linewidth]
    \scriptsize 

    \textbf{\# Role}

    You are a professional Instruction Violation Sample Generation Expert. 

    \textbf{\#\# Task}
    
    Your task is to generate an "Imperfect Response" that violates specific instruction mandates based on a user query and a corresponding "Good Response" that fully complies with the instructions.

    \textbf{\#\# Core Generation Principles}

    1. Content Retention: The content of the synthesized Imperfect Response should be derived from the Good Response, maintaining the primary information, resource citations, and key arguments.
    
    2. Targeted Violation: Specifically violate the corresponding requirements within the instructions based on the designated violation type.
    
    3. Naturalness: Violations should be natural and not overtly forced, mimicking common errors models make when attempting to follow complex instructions.
    
    4. Single Violation: Violate only one primary compliance dimension at a time to ensure the bad sample clearly corresponds to a specific violation type.
    
    5. Scenario Alignment: Violations must be based on the specific requirements of the current business scenario and should not apply rules from unrelated contexts.

    \textbf{\#\# Violation Types by Dimension}
    
    1. Role Compliance Violation: Identity Misalignment, Capability Overreach, Boundary Handling Failure, Organizational Inconsistency, Recommendation Rigidity.

    2. Process Compliance Violation: Sequence Error, Mandatory Step Omission.

    3. Format Compliance Violation: Tag Usage Integrity Failure, Prohibited Format Usage, Structural Non-alignment, Markdown Non-standardization, Element Ordering Error.

    4. Content Quality Violation: Factuality and Accuracy Failure, Information Filtering Non-compliance, Organizational Irregularity, Depth and Richness Deficit.

    5. Constraint Compliance Violation: Prohibited Constraint Breach, Safety and Regulatory Violation.

    6. Style Compliance Violation: Tone Inconsistency, Vocabulary Non-standardization, Expressive Misalignment, Persona Feature Omission, Linguistic Quality Issues, Expression Misalignment.

    \textbf{\# Output Format}
    \begin{verbatim}
    
    <think>
    [Thought process content]
    </think>
    
    <answer>
    [Actual response content]
    </answer>
    
    <violation_detail>
    Specifically violated the following mandates: 
    1. [Instruction 1]; 2. [Instruction 2]...
    </violation_detail>
    
    \end{verbatim}
    
    \end{tcolorbox}
    \caption{Prompt for Instruction Following Data Augmentation}
    \label{fig:prompt_IF_syn}
\end{figure}

\begin{figure}[H]
    \centering

    \begin{tcolorbox}[
    colback=gray!5!white,
    colbacktitle=gray!30!white,
    colframe=black,
    boxrule=1pt,
    left=1mm, right=1mm,
    halign title=center,
    rounded corners,
    coltitle=black,
    title=Prompt for Instruction Following Assessment, width=\linewidth]
    \scriptsize 
    \textbf{\# Role}
    
    You are a professional Instruction Following Assessment Expert, responsible for evaluating instruction compliance.
    
    \textbf{\#\# Task}
    
    Evaluate adherence of model responses based on provided instructions.
    
    \textbf{\# Evaluation Criteria}
    
    \textbf{\#\# Evaluation Dimensions}   
    
    \textbf{\#\#\# Role Compliance}
    
    - Identity Fidelity: Assess whether the response consistently reflects the defined persona.
    
    - Scope Adherence: Verify that the model acts strictly within the defined scope.
    
    - Refusal Policy Alignment: Evaluate the appropriateness of handling out-of-scope requests, ensuring that refusals or redirections align with the specified interaction policy.
    
    - Organizational Alignment: Ensure strict compliance with brand, platform, or organizational constraints.
    
    - Contextual Flexibility: Verify direct answering without forced recommendations for general queries.

    \textbf{\#\#\# Process Compliance}
    
    - Sequence Correctness: Verify the model's adherence to the mandated execution order specified in the instructions.
    
    - Mandatory Step Integrity: Evaluate whether all required intermediate steps and essential components have been fully executed without omission.
    
    \textbf{\#\#\# Format Compliance}
    
    - Tag Usage Integrity: Verify the correct application and syntax of mandatory tags.
    
    - Prohibited Format Avoidance: Ensure the strict avoidance of forbidden structural patterns or restricted formatting elements.
    
    - Structural Alignment: Evaluate the model's adherence to specified output schemas and predefined content structures.
    
    - Markdown Format Adherence: Assess compliance with standard Markdown conventions and specific formatting constraints. 
    
    - Element Ordering Correctness: Verify the correct ordering of format elements as mandated by the instructions.

    \textbf{\#\#\# Content Quality Compliance}
    
    - Factuality and Accuracy: Verify the objective truthfulness of the information, ensuring the avoidance of hallucinations.
    
    - Information Filtering: Assess whether the model correctly filters content according to the exclusion/inclusion criteria specified.
    
    - Structural Organization: Verify whether the content arrangement strictly aligns with instructions and evaluate the logical coherence of the response.
    
    - Depth and Richness: Evaluate whether the content satisfy the prescribed requirements for richness and depth.
    
    \textbf{\#\#\# Constraint Compliance}
    
    - Prohibited Constraint Adherence: Verify strict compliance with all items explicitly marked as "prohibited," ensuring no such items appear in the response.

    - Safety and Regulatory Compliance: Evaluate the model's adherence to safety protocols and regulatory standards.
    
    \textbf{\#\#\# Style Compliance}
    
    - Tone Consistency: Verify adherence to the mandated tone.

    - Vocabulary Standards: Evaluate word choices, ensuring use of professional terms and avoidance of forbidden language.

    - Expressive Alignment: Assess whether the expression style meets instruction criteria.

    - Persona Embodiment: Verify the inclusion of specific traits.

    - Linguistic Quality: Evaluate grammar, fluency, and logical consistency while avoiding redundancy.

    - Expression Adherence: Ensure compliance with specific phrasing mandates or restricted expression constraints.
    
    \textbf{\#\# Scoring} 
    
    5=Fully compliant; 4=Minor deviations; 3=Notable deviations; 2=Clearly non-compliant; 1=Severe violation.

    \textbf{\# Output Format}  
    
    \begin{verbatim}
    {
      "instruction_following_score": {
        "overall_score": 1-5,
        "dimension_scores": {
          "role_compliance": 1-5 or "N/A",
          "process_compliance": 1-5 or "N/A",
          "format_compliance": 1-5 or "N/A",
          "content_quality_compliance": 1-5 or "N/A",
          "constraint_compliance": 1-5 or "N/A",
          "style_compliance": 1-5 or "N/A"
        },
        "detailed_reasoning": {
          "role_compliance": "Detailed justification.",
          "process_compliance": "Detailed justification.",
          "format_compliance": "Detailed justification.",
          "content_quality_compliance": "Detailed justification.",
          "constraint_compliance": "Detailed justification.",
          "style_compliance": "Detailed justification."
        },
        "critical_violations": [
          "List all critical violations. Empty if none."
        ],
        "summary": "Comprehensive assessment summary."
      }
    }
    \end{verbatim}
    
    \end{tcolorbox}
    \caption{Prompt for Instruction Following Assessment}
    \label{fig:prompt_IF}
\end{figure}
\clearpage

\begin{figure}[!htbp]
    \centering

    \begin{tcolorbox}[
    colback=gray!5!white,
    colbacktitle=gray!30!white,
    colframe=black,
    boxrule=1pt,
    left=1mm, right=1mm,
    halign title=center,
    rounded corners,
    coltitle=black,
    title=Prompt for Factual Consistency Assessment, width=\linewidth]
    \scriptsize 

    \textbf{\# Role}

    You are a professional Factual Error Verification Expert responsible for identifying hallucination-related factual errors in the agent responses.
    
    \textbf{\#\# Task}
    According to the received complete conversation, including the user query, the tools called by the system and their return results, and the final response, check the factual errors exist in the response.

    \textbf{\# Input Format}
    
    \begin{verbatim}
    {
      "tool_responses": Content of tool responses,
      "assistant_response": The final response from the assistant,
      "id_list": A list of IDs to be verified,
      "location": User's current location at the time of query,
      "current_time": Time of the query
    }
    \end{verbatim}
    
    \textbf{\# Evaluation Criteria} 
    
    Strictly follow the order of id\_list in the input to verify each ID one by one according to the following evaluation steps:
    
    \textbf{\#\# Item Evaluation Steps}
    
    - Step 1: Information Extraction and Preparation. Extract the information corresponding to the ID in the tool responses (if the ID does not exist in the tool\_responses, it is a hallucination); Extract opening summary in the assistant\_response applied to every ID; Extract the name and text description associated with the ID in the assistant\_response, and combine it with the opening summary to obtain the text info to be verified.
    
    - Step 2: Key Information Verification. Verify if the following information matches tool\_responses: Merchant/Product Name, Address, Distance, Rating, Price, Service Scenarios, Service Support, Category.
    
    Step 3: Severity Judgment. Determine if the error is serious (significantly affects user decision-making). Minor errors should not be output.

    \textbf{\#\# Content Evaluation}
    Identify non-ID-specific content (background knowledge, general guides, etc.) and verify if there are obvious factual errors or misleading statements.
    
    \textbf{\# Output Format}
    \begin{verbatim}
    {
      "Item Hallucination": [
        {
          "ID Name 1": "Name of ID 1",
          "Questions": [
            {
              "Question 1": "Description of the specific problem and the basis for it"
            },
            {
              "Question 2": "..."
            }
          ]
        },
        {
          "ID Name 2": "Name of ID 2",
          "Questions": [...]
        }
      ],
      "Content Hallucination": [
        {
          "Question 1": "Factual error description in extra content"
        },
        {
          "Question 2": "Misleading statement description in extra content"
        }
      ]
    }
    \end{verbatim}
    
    \end{tcolorbox}
    \caption{Prompt for Factual Consistency Assessment}
    \label{fig:prompt_hallucination}
\end{figure}

\begin{figure}[H]
    \centering

    \begin{tcolorbox}[
    colback=gray!5!white,
    colbacktitle=gray!30!white,
    colframe=black,
    boxrule=1pt,
    left=1mm, right=1mm,
    halign title=center,
    rounded corners,
    coltitle=black,
    title=Prompt for Query-Item Relevance Assessment, width=\linewidth]
    \scriptsize 

    \textbf{\# Role}

    You are a professional Relevance Evaluation Expert, specifically responsible for assessing the relevance between recommended merchants/products and user queries.
    
    \textbf{\#\# Task}
    Analyze merchant or product information to determine its relevance based on both the user's original query and the rewritten query.

    \textbf{\# Input Format}
    \begin{verbatim}
    {
      "query": Original search query from user,
      "rewritten query": Rewritten query by the system,
      "output": Merchant or product information
    }
    \end{verbatim}

    \textbf{\# Evaluation Criteria}
    
    \textbf{\#\# Scoring}

    - Relevant (score=2): Fully relevant. Core elements such as brand, category, and product are completely matched.
    
    - Weakly Relevant (score=1): Partially relevant. Similar categories or only partial elements matched.
    
    - Irrelevant (score=0): Completely irrelevant. Significant differences in category or core elements do not match.

    \textbf{\#\# Core Principles}
    
    1. No Speculation: Absolutely no inferring or assuming information that is not explicitly present in the query and model output.
    
    2. Brand Match: Verify the brand in the merchant name first; if not found, check associated products or sub-items. Do not infer the merchant's brand solely from the brands of its sub-items.
    
    3. Numerical Standards:
    - High Rating: $\ge$ 4.0
    - Nearby: Relative distance $\le$ 5 km
    - Short Delivery: $\le$ 40 minutes
    - Budget: Permits a tolerance of $±$10$\%$
    
    \textbf{\#\# Evaluation Steps}
    
    - Step 1: Identify Category and Query Type. Recognize the core category of the query.
    
    - Step 2: Extract Query Elements. Identify all evaluation elements within the query, including:
    
        **Core Elements**: Brand, merchant name, product name.
        
        **Detailed Conditions**: Rating requirements, budget constraints, delivery speed, business hours.
        
        **Location Constraints**: Specific position names or proximity indicators.
        
    - Step 3: Verify Category Consistency. Category consistency serves as the foundational criterion for relevance assessment. If the category is inconsistent, the result must be labeled Irrelevant, and no further checks are required. If the category is consistent, proceed to evaluate the remaining elements.
    
    - Step 4: Verify Core Element Consistency. Check whether all core elements identified in the query are present in the model output. If any core element is missing, the result should be labeled Weakly Relevant.
    
    - Step 5: Verify Location Compliance. Confirm that the model output satisfies all location requirements specified in the query. This includes compliance with single or multiple specified positions as well as any nearby distance conditions.
    
    - Step 6: Verify Numerical and Detailed Conditions. Evaluate whether the model output meets any additional numerical or conditional requirements present in the query, such as rating thresholds, price limits, business hours, or delivery time.
    
    - Step 7: Determine Final Relevance Level. Based on the preceding checks: If the category is consistent but other conditions are not fully satisfied, the result is Weakly Relevant. If the category is consistent and all required conditions are satisfied, the result is Fully Relevant. If the category is inconsistent, the result is Irrelevant, regardless of any other conditions.

    \textbf{\# Output Format}
    \begin{verbatim}
    {
      "Relevance": {
        "Reason": "Brief explanation of judgment",
        "Score": 0 or 1 or 2
      }
    }
    \end{verbatim}
    
    \end{tcolorbox}
    \caption{Prompt for Query-Item Relevance Assessment}
    \label{fig:prompt_relev}
\end{figure}

\begin{figure}[H]
    \centering

    \begin{tcolorbox}[
    colback=gray!5!white,
    colbacktitle=gray!30!white,
    colframe=black,
    boxrule=1pt,
    left=1mm, right=1mm,
    halign title=center,
    rounded corners,
    coltitle=black,
    title=Prompt for Item Ranking Assessment, width=\linewidth]
    \scriptsize 

    \textbf{\# Role}
    
    You are a recommendation purchase prediction model for an e-commerce platform. Your task is to predict whether a user will purchase the recommended product based on the user's personal profile, search query, and the list of recommended products. 
    
    \textbf{\# Input Information}   
    
    \underline{User Profile}: basic information (e.g., location, age range, gender, date of birth, adult status, zodiac sign, occupation, vehicle ownership, marital status, parental status), consumer profile (e.g., purchasing power, lifestyle refinement, audience segment, secondary audience segment), preference (e.g., behavioral preferences, product preferences), and real-time context (e.g., current time, current location, current occasion, current situation).

    \underline{Query}: current query, current location, current date.
    
    \underline{Recommended List}: product name, associated merchant name, relative location, delivery threshold, delivery fee, delivery time, sales volume, price.
    
    \textbf{\# Task}
    
    Predict whether the user will purchase this recommended product.
    
    \textbf{\# Output Format}
    
    Your output must be one of the following two words only:
    
    \textcolor{darkgreen}{"yes"}  - if you predict the user will purchase the recommended product.
    
    \textcolor{red}{"no"} - if you predict the user will not purchase the recommended product.
    
    \textbf{\# Prediction Guidelines}
    
     - Match the recommended product with the user's current targeting of scenarios.
    
     - Consider the user's behavioral preferences and product preferences.
    
     - Evaluate whether the price is reasonable (compared to the user's purchasing power).
    
     - Consider relative location, delivery fee, and delivery time.
    
     - Check if the product category aligns with the user's interests.
    
     - Evaluate merchant reputation (sales volume).
    
     - Consider the match between user profile (age range, occupation type, etc.) and the product.
    
    \textbf{\# Important Rules}
    
    1. Only output \textcolor{darkgreen}{"yes"} or \textcolor{red}{"no"}.
    
    2. Do not provide any explanations or additional text.
    
    3. Do not add punctuation.
    
    4. Do not add line breaks.
    
    \end{tcolorbox}
    \caption{Prompt for Item Ranking Assessment}
    \label{fig:prompt_item_ranking}
\end{figure}

\begin{figure}[H]
    \centering

    \begin{tcolorbox}[
    colback=gray!5!white,
    colbacktitle=gray!30!white,
    colframe=black,
    boxrule=1pt,
    left=1mm, right=1mm,
    halign title=center,
    rounded corners,
    coltitle=black,
    title=Prompt for Behavior Prediction Assessment, width=\linewidth]
    \scriptsize 

    \textbf{\# Role}
    
    You are a model predicting whether a user is interested in the recommended list content. Judging whether a user is interested in the recommended content depends on whether the user will take further action, such as clicking or placing an order. The input consists of three parts:
    
    \underline{User Profile}: basic information (e.g., location, age range, gender, date of birth, adult status, zodiac sign, occupation, vehicle ownership, marital status, parental status), consumer profile (e.g., purchasing power, lifestyle refinement, audience segment, secondary audience segment), preference (e.g., behavioral preferences, product preferences), and real-time context (e.g., current time, current location, current occasion, current situation).
    
    \underline{Query}: current location, current time, current query.
    
    \underline{Recommended List}: merchants (name, review score, business hours, etc.), products (associated merchant, price, relative location, review score, etc.).
    
    \textbf{\# Evaluation Steps}
    
    Step 1. Hard Constraints Filtering: e.g., no car ownership→do not recommend car-related products.
    
    Step 2. Determine whether the user has deep engagement needs.
    
    Step 3. Intent/Scene Judgment.
    
    Step 4. Preference and Level Adjustment.
    
    \textbf{\# Output Requirements (Must be strictly followed)}
    
    - The output can only be \textcolor{darkgreen}{"1"} or \textcolor{red}{"0"}.
    
    - Do not output any explanations or other characters.

    \end{tcolorbox}
    \caption{Prompt for Behavior Prediction Assessment}
    \label{fig:prompt_behavior}
\end{figure}

\end{document}